\documentclass[12pt]{article}

\usepackage{epsfig}
\usepackage{amssymb,amsmath}

\newcommand{\bea}{\begin{eqnarray}}
\newcommand{\eea}{\end{eqnarray}}

 \makeatletter
\def\alt{\mathrel{\mathpalette\gl@align<}}
\def\agt{\mathrel{\mathpalette\gl@align>}}
\def\gl@align#1#2{\lower.6ex\vbox{\baselineskip\z@skip\lineskip\z@
\ialign{$\m@th#1\hfil##\hfil$\crcr#2\crcr\sim\crcr}}} \makeatother

\begin{document}
\begin{flushright}
BA-07-23
\end{flushright}
\vspace*{1.0cm}

\begin{center}
\baselineskip 20pt {\Large\bf Higgs Boson Mass From Gauge-Higgs
Unification
 } \vspace{1cm}

{\large Ilia Gogoladze$^{a,}$\footnote{ E-mail:
ilia@physics.udel.edu}, Nobuchika Okada$^{b,c,}$\footnote{ E-mail:
okadan@post.kek.jp} and Qaisar Shafi$^{a,}$\footnote{ E-mail:
shafi@bartol.udel.edu} } \vspace{.5cm}

{\baselineskip 20pt \it
$^a$Bartol Research Institute, Department of Physics and Astronomy, \\
University of Delaware, Newark, DE 19716, USA \\
\vspace{2mm} $^b$Department of Physics, University of Maryland,
College Park,  MD 20742, USA \\
\vspace{2mm} $^c$Theory Division, KEK, Tsukuba 305-0801, Japan }
\vspace{.5cm}

\vspace{1.5cm} {\bf Abstract}
\end{center}

In certain five dimensional gauge theories
 the Standard Model Higgs doublet is identified,
 after compactification on the orbifold $S^1/Z_2$,
 with the zero mode of the fifth component of the gauge field.
An effective potential for the Higgs field is generated
 via quantum corrections, triggered by the breaking
 of the underlying gauge symmetry through boundary conditions.
The quartic Higgs coupling can be estimated at low energies
 by employing the boundary condition that it vanishes
 at the compactification scale $\Lambda$, as
 required by five dimensional gauge invariance.
For $\Lambda \gtrsim 10^{13}-10^{14}$ GeV,
 the Standard Model  Higgs boson mass is found to be $m_H = 125 \pm$ 4 GeV,
 corresponding to a top quark pole mass
 $M_t = 170.9 \pm 1.8$ GeV.
 A more complete (gauge-Higgs-Yukawa) unification
can be realized for $\Lambda \sim 10^8$ GeV, which happens to be the
scale at which the SU(2) weak coupling and the top quark Yukawa
coupling have the same value. For this case, $m_H = 117\pm4$ GeV.

\thispagestyle{empty}

\newpage

\addtocounter{page}{-1}

\baselineskip 18pt

It seems reasonable to hope that the Standard Model (SM) Higgs boson
will likely be found in the near future, most likely at the LHC. The
discovery should reveal a great deal about the origin of electroweak
breaking and the mechanism responsible for generating the quark and
charged lepton masses. A precise measurement of the Higgs mass $m_H$
should help distinguish between various competing theoretical ideas.
One could argue, for example, that the MSSM would be one of the
favored schemes if $m_H$ turns out to be close to its current
experimental lower limit of 114.4 GeV \cite{LEP2}. However, values
of $m_H$ around 125 GeV or larger, would suggest a much more serious
consideration of other competing ideas. For instance, in a class of
higher dimensional supersymmetric orbifold models in which the 4D
N=1 supersymmetry is broken at $M_\text{GUT}$, the Higgs mass $m_H = 145
(\pm 19)$ GeV \cite{NCY1}. The SM gauge couplings in these models
are unified at $M_\text{GUT}$ by employing a non-canonical U(1)$_Y$. An
important extension of these ideas implements gauge and Yukawa
coupling unification at $M_\text{GUT}$ \cite{NCU}. For instance,
with gauge-top quark Yukawa coupling unification and with SUSY
broken at $M_{\text{GUT}}$, the SM Higgs boson mass turns out to be
$135\pm 6$ GeV \cite{NCU}. Somewhat larger values for the Higgs
mass, $144\pm 4$ GeV, are found with gauge-bottom quark Yukawa
coupling unification \cite{NCU}.

 In this letter we attempt to
estimate $m_H$ by employing the idea of gauge-Higgs unification
(GHU) which has attracted a fair amount of recent attention
\cite{GHU1}-\cite{GHY}. We consider, in particular, 5D models
compactified on an orbifold $S^1/Z_2$, such that the zero mode of
the fifth component of the bulk gauge field can be identified with
the SM Higgs doublet. The so-called "gauge-Higgs" condition , to be
explained shortly, enables us to estimate the SM Higgs mass $m_H$.
Using two-loop renormalization group equations (RGEs), we find that
$m_H$ exceeds the LEP2 lower bound if the compactification scale
$\Lambda \gtrsim 10^6$ GeV. The weak SU(2) gauge coupling and the
top Yukawa coupling have the same magnitude at scales close to
$10^8$ GeV. If the latter is identified with the compactification
scale, the Higgs mass $m_H$ is predicted to be $117\pm 4$ GeV.
Finally, following \cite{NCY1, NCU}, if $\Lambda$ is identified with
the SM gauge coupling unification scale of order $4\times 10^{16}$
GeV which is possible with non-canonical U(1)$_Y$, $m_H = 125\pm 4$
GeV.

We consider 5D Gauge-Higgs Unification (GHU) model
with the fifth dimension compactified on the orbifold $S^1/Z_2$ which
yields  a chiral "low energy" theory in  four dimensions. In GHU
models, the 5D  bulk gauge symmetry is broken
 down to the SM by   imposing suitable
 boundary conditions.
 The SM Higgs doublet
 emerges as a zero-mode of the fifth component of
 the higher dimensional gauge field.
The higher dimensional gauge symmetry prevents the appearance of a
tree level scalar potential. However, since the bulk gauge symmetry
is broken by the boundary condition, a quartic Higgs potential is
induced through quantum correction. In particular, at one-loop
level,
 the effective Higgs potential
 has been found to be finite \cite{GHU-finiteness}.
This finiteness can be interpreted as a  remnant
 of the higher dimensional gauge invariance
 and its non-local breaking by the boundary condition.
The mass scale of the finite effective potential
 is controlled  by the Kaluza-Klein (KK) mass, $1/R$,
 where $R$ is the radius of the fifth dimension.

Recently, a new phenomenological treatment of GHU models
 has been proposed \cite{GHUcond}.
It has been  shown that the effective  SM Higgs quartic coupling
$\lambda$
 calculated in a given GHU model coincides with the one
 radiatively generated in the effective low energy  theory
 (without a  quartic coupling at tree level), provide
  the compactification scale $\Lambda=1/(2 \pi R)$
 is identified with the cutoff scale in evaluating  quantum corrections.
This  implies that the higher dimensional gauge invariance
 is  restored at  scales  smaller than the extra dimensional
 volume $2 \pi R$, such that  there is  no effective quartic Higgs coupling
 at high energy, it appears, should be applicable to any GHU model. Thus,
  using renormalization group equations (RGEs),  we can evaluate the SM quartic
Higgs coupling by requiring that  $\lambda$ vanishes at
$\Lambda=1/(2 \pi R)$. This boundary condition for $\lambda$ is
called \cite{GHUcond} the ``gauge-Higgs condition''. It is
reminiscent of the so-called  vacuum stability bound on $m_H$
\cite{stability1}.


Corrections to the Higgs mass squared in GHU models, on the other
hand, are very much dependent
 on the  particle content and imposed boundary conditions (see, for instance  \cite{GHUcond}), and so in
 this paper we can treat the Higgs mass$^2$  as a free parameter
 in the low energy effective theory, to be suitably adjusted to
 yield the desired electroweak symmetry breaking.

In contrast, as we  previously  discussed,
 the gauge-Higgs condition is a model-independent condition
 imposed on the low energy effective theory.
Therefore, when applied   to
 the SM Higgs doublet,
 we can obtain a prediction for the physical Higgs boson mass
 as a function of the compactification scale.
The Higgs boson mass prediction we will obtain is magnitude wise
 the same as the vacuum stability  bound mentioned above.
However, we have a physics interpretation for it, namely  the GHU
model(s) provides an  ultraviolet completion of the SM. In the
bottom-up picture, the quartic Higgs coupling in the SM
 reaches zero at the compactification scale,
 at which point  the GHU model takes over.
The Higgs potential is smoothed away
 such that no instability in the Higgs potential occurs.

We are now ready to discuss the physical  Higgs boson mass $m_H$ in
this setup. With  the electroweak symmetry breaking property
implemented, $m_H$  is determined by its quartic coupling. Imposing
the ``gauge-Higgs condition'' for the Higgs quartic coupling
 at a given compactification scale $\Lambda(=1/(2 \pi R))$
 and solving the two loop RGEs, \cite{RGE}, towards the electroweak scale,
 we obtain the Higgs boson mass as
 a function of  $\Lambda$. Namely,
\bea
 m_H (m_H) = \sqrt{\lambda(m_H)} \; v .
\eea

For the  three SM gauge couplings, we have
\bea
 \frac{d g_i}{d \ln \mu} =  \frac{b_i}{16 \pi^2} g_i^3 +\frac{g_i^3}{(16\pi^2)^2}
\sum_{j=1}^3B_{ij}g_j^2, \label{gauge} \eea
 where $ \mu$ is the renormalization scale,
$g_i$ ($i=1,2,3$) are the SM  gauge couplings  and
\begin{equation}
b_i = \left(\frac{41}{10},-\frac{19}{6},-7\right),~~~~~~~~
 { b_{ij}} =
 \left(
  \begin{array}{ccc}
  \frac{199}{50}& \frac{27}{10}&\frac{44}{5}\\
 \frac{9}{10} & \frac{35}{6}&12 \\
 \frac{11}{10}&\frac{9}{2}&-26
  \end{array}
 \right).
\end{equation}
The top quark pole mass is taken to be
 $M_t= 170.9 \pm 1.8$ GeV,  \cite{Tevatron},
with
 $(\alpha_1, \alpha_2, \alpha_3)=(0.01681, 0.03354, 0.1176)$
 at $M_Z$ \cite{PDG}.
For the top Yukawa coupling $y_t$, \cite{RGE},
\bea \label{ty}
 \frac{d y_t}{d \ln \mu}
 = y_t  \left(
 \frac{1}{16 \pi^2} \beta_t^{(1)} + \frac{1}{(16 \pi^2)^2} \beta_t^{(2)}
 \right).
\label{topYukawa}
\eea
Here the one-loop contribution is
\bea
 \beta_t^{(1)} =  \frac{9}{2} y_t^2 -
  \left(
    \frac{17}{20} g_1^2 + \frac{9}{4} g_2^2 + 8 g_3^2
  \right),
\label{topYukawa-1}
\eea
while the two-loop contribution is given by
\bea
\beta_t^{(2)} &=&
 -12 y_t^4 +   \left(
    \frac{393}{80} g_1^2 + \frac{225}{16} g_2^2  + 36 g_3^2
   \right)  y_t^2  \nonumber \\
 &+& \frac{1187}{600} g_1^4 - \frac{9}{20} g_1^2 g_2^2 +
  \frac{19}{15} g_1^2 g_3^2
  - \frac{23}{4}  g_2^4  + 9  g_2^2 g_3^2  - 108 g_3^4 \nonumber \\
 &+& \frac{3}{2} \lambda^2 - 6 \lambda y_t^2 .
\label{topYukawa-2}
\eea
In solving Eq.(\ref{ty})  from $M_t$ to the compactification scale
$\Lambda$,
 the initial top Yukawa coupling at $\mu=M_t$
 is determined from the relation
 between the pole mass and the running Yukawa coupling \cite{RGE1},
\bea
   \frac{M_t}{m_t(M_t)} \simeq
  1 + \frac{4}{3} \frac{\alpha_3(M_t)}{\pi}
  + 10.91  \left( \frac{\alpha_3(M_t)}{\pi} \right)^2,
\eea
 with $ y_t(M_t) = \sqrt{2} m_t(M_t)/v$, where $v=246.2$ GeV.

The RGE  for the Higgs quartic coupling is given by \cite{RGE},

\bea
\frac{d \lambda}{d \ln \mu}
 =   \frac{1}{16 \pi^2} \beta_\lambda^{(1)}
   + \frac{1}{(16 \pi^2)^2}  \beta_\lambda^{(2)},
\label{self}
\eea
with
 \bea
 \beta_\lambda^{(1)}=
 12 \lambda^2 -
 \left(  \frac{9}{5} g_1^2+9 g_2^2  \right) \lambda
 + \frac{9}{4}
 \left(
 \frac{3}{25} g_1^4 + \frac{2}{5} g_1^2 g_2^2 +g_2^4
 \right)
 + 12 y_t^2 \lambda -12 y_t^4 ,
\label{self-1}
\eea
and
\bea
  \beta_\lambda^{(2)} &=&
 -78 \lambda^3  + 18 \left( \frac{3}{5} g_1^2 + 3 g_2^2 \right) \lambda^2
 - \left( \frac{73}{8} g_2^4  - \frac{117}{20} g_1^2 g_2^2
 + \frac{2661}{100} g_1^4  \right) \lambda - 3 \lambda y_t^4
 \nonumber \\
 &+& \frac{305}{8} g_2^6 - \frac{289}{40} g_1^2 g_2^4
 - \frac{1677}{200} g_1^4 g_2^2 - \frac{3411}{1000} g_1^6
 - 64 g_3^2 y_t^4 - \frac{16}{5} g_1^2 y_t^4
 - \frac{9}{2} g_2^4 y_t^2
 \nonumber \\
 &+& 10 \lambda \left(
  \frac{17}{20} g_1^2 + \frac{9}{4} g_2^2 + 8 g_3^2 \right) y_t^2
 -\frac{3}{5} g_1^2 \left(\frac{57}{10} g_1^2 - 21 g_2^2 \right)
  y_t^2  - 72 \lambda^2 y_t^2  + 60 y_t^6.
\label{self-2}
\eea

In Figure~1, we plot the  Higgs boson mass $m_H$
 as a function of the compactification scale
 for an input top quark pole mass $M_t= 170.9 \pm 1.8$ GeV.
The current Higgs boson mass bound,
 $m_H > 114.4$ GeV, from  LEP2 \cite{LEP2} is displayed as horizontal line.
Requiring  the compactification scale $\Lambda \lesssim M_{Pl}=1.2
\times 10^{19}$ GeV,
 the upper bound on the Higgs boson mass is found to be
 $m_H \lesssim 129$ GeV.
A compactification scale larger than $2.5 \times 10^6$ GeV results
in a Higgs boson mass which is consistent with the current lower
bound. Higgs boson masses for varying top quark pole mass
 and the compactification scale are listed in Table~1.

%

\begin{figure}[t, width=12cm, height=8cm]
\begin{center}
{\includegraphics[height=7cm]{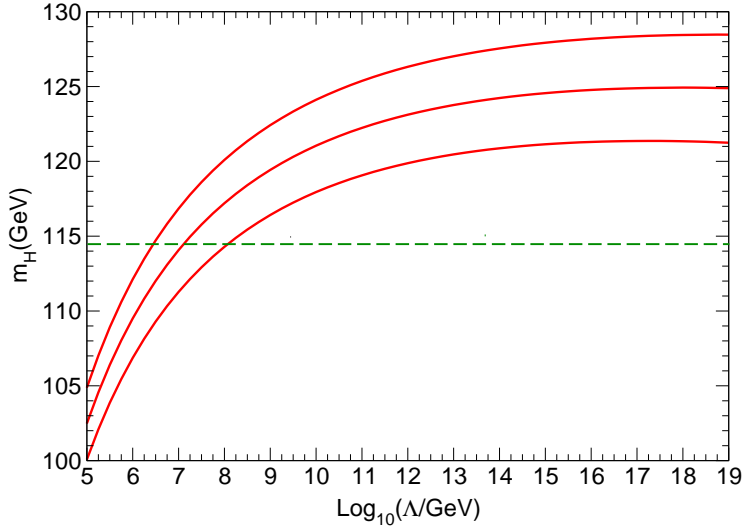} }
\end{center}
\caption{ \small Higgs boson mass prediction versus
 the compactification scale, $\mbox{Log}_{10}( \Lambda/\mbox{GeV})$.
The solid lines (in red) correspond from bottom to top to input top
quark pole masses,
 $M_t=$169.1, 170.9 and 172.8 GeV.
The horizontal line shows the current Higgs mass bound,
 $m_H \geq 114.4$ GeV, from LEP2.
}
\vspace{-1.5cm}
\end{figure}
\begin{center}
\begin{table}[t]
\centering
\begin{tabular}{c|ccc}
$\Lambda$ [GeV]  &~~ $M_t=169.1$ GeV ~~
                       &~~ $M_t=170.9$ GeV ~~
                       &~~ $M_t=172.8$ GeV ~~\\
\hline \hline
$10^5$                 & 100.0  &  102.4 & 104.9 \\

$10^7$                 & 111.3  &  114.1 & 116.9 \\

$10^9$                 & 116.4  &  119.4 & 122.5 \\

$10^{11}$              & 119.1  &  122.2 & 125.4 \\

$10^{13}$              & 120.5  &  123.8 & 127.1 \\

$10^{15}$              & 121.1  &  124.6 & 128.0 \\

$10^{17}$              & 121.4  &  124.9 & 128.4 \\

$10^{19}$              & 121.2  &  124.9 & 128.5 \\
\end{tabular}
\caption{\small Higgs boson masses (in GeV) for varying $M_t$ and
$\Lambda$. }
\end{table}
\end{center}

In the (simplest) GHU model, the unification of gauge and Yukawa
interactions would imply fermion mass coincide with the weak gauge
boson mass at  $\Lambda$. With the possible exception of the top
quark, as we will see, this is clearly not acceptable. To realize
the hierarchy of fermion masses in the SM, more elaborate GHU model
must  be considered. There have been various efforts along  this
direction \cite{Csaki}. As far as the top quark is concerned the
running top Yukawa coupling and the SU(2) gauge coupling  meet at a
intermediate  scale (see Figure~2), and it would be natural to
identify this 'merger' point with the compactification scale. This
observation allows us to realize gauge-Higgs and gauge-top Yukawa
coupling unification at $\Lambda$, \cite{GHY}, in this case more
precise  prediction for the Higgs boson mass is obtained, namely
$m_H=117\pm 4$ GeV (see Table 2).

%
\begin{figure}[bt,width=12cm, height=8cm]
\begin{center}
{\includegraphics[height=7cm]{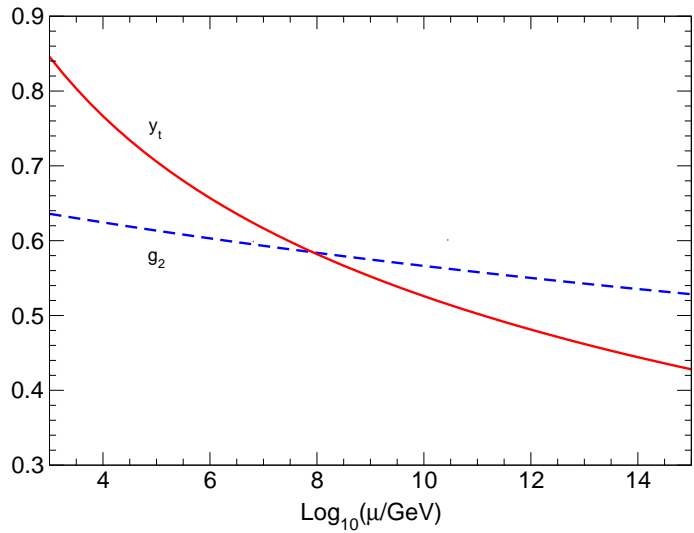} }
\end{center}
\caption{ \small Plot of SU(2) gauge  $g_2(\mu)$ (dashed line) and
top Yukawa $y_t(\mu)$ (solid line) coupling versus $\mbox{Log}_{10}(
\mu /\mbox{GeV})$. The two couplings  merge at $\mu=8.41 \times
10^7$ GeV,  for $M_t=170.9$ GeV.}
\vspace{-1.5cm}
\end{figure}
\begin{center}
\begin{table}[b]
\centering
\begin{tabular}{c|ccc}
                       &~~ $M_t=169.1$ GeV ~~
                       &~~ $M_t=170.9$ GeV ~~
                       &~~ $M_t=172.8$ GeV ~~\\
\hline \hline
$\Lambda$ [GeV]  & $ 3.26 \times 10^7 $
                       & $ 8.41 \times 10^7 $
                       & $ 2.34 \times 10^8 $ \\
$m_H$ [GeV]            & 112.9  &  117.0 & 121.1
\end{tabular}
\caption{ \small Higgs boson mass (in GeV) for varying $M_t$ with
$g_2(\Lambda)=y_t(\Lambda)$  at $\mu=\Lambda$ (see Figure 2).}
\end{table}
\end{center}

Another interesting possibility is to identify $\Lambda$ with
$M_\text{GUT} \sim 10^{14}-10^{16}$ GeV. In this case, $m_H=125\pm
4$ GeV. Although the three SM gauge couplings do not meet at
$M_\text{GUT}$ with a canonical normalization of $5/3$  for
U(1)$_Y$, a different choice, say $4/3$, can lead to gauge coupling
unification at $M_\text{GUT} = 4 \times 10^{16}$ GeV \cite{NCY1,
NCU}. Note that the upper bound on $m_H$ we have found for GHU
models is well below the bound of 180 GeV or so, obtained from the
entirely  different requirement that $\lambda$ should remain
perturbative between $M_Z$ and $M_{Pl}$.

In conclusion, we have considered  gauge-Higgs unification
 in five dimensions with  $S^1/Z_2$ orbifold compactification,
 such that  the SM Higgs doublet  emerges
 as the fifth component of the higher dimensional gauge field.
Due to  higher dimensional gauge invariance,
 there is no tree level Higgs potential in the
  effective four dimensional theory. This is modified by
  quantum corrections which generate  a quartic Higgs potential
 with an effective  finite  cutoff,
 $\Lambda=1/(2 \pi R)$.
While the induced Higgs self energy is highly dependent
 on the details of the model considered, the quartic Higgs coupling
 coincides with the one obtained in the low energy massless
 theory by employing the gauge-Higgs condition at $\Lambda$.
Imposing this condition on the Higgs quartic coupling
 in the SM and employing two loop RGEs,
 we have obtained  predictions for
 the Higgs boson mass as a function of the compactification scale.

Finally,  a comment on cosmology is in order here. The existence of
 (non-baryonic) dark matter
 has been established from various observations of the present universe.
Except for the Higgs sector, the gauge-Higgs unification model
 shares the same structure
 as the Universal Extra Dimension model \cite{UED,Antoniadis:1990ew},
 so that the first KK excitation  of U(1)$_Y$ gauge boson is a plausible dark matter candidate \cite{UED-DM}.
In our case, the mass  of this  particle can be very high and,
indeed, it can  exceed the unitarity limit on the mass of cold dark
matter as a thermal relic. However, there are   various
possibilities  to realize  superheavy  dark matter, such as its
production through  inflaton decay.

\section*{Acknowledgments}

We  thank Y. Mimura,  M.~ur Rehman and V.~N.~Senoguz for very
helpful  discussions.
 N.O. would like to thank the Particle
Physics Group
 of the University of Delaware for the hospitality
 during his visit.
This work is supported in part by
 the DOE Grant \# DE-FG02-91ER40626 (I.G. and Q.S.),
 and
 the Grant-in-Aid for Scientific Research from the Ministry
 of Education, Science and Culture of Japan,
 \#18740170 (N.O.).


\end{document}